\documentclass[twocolumn,showpacs,preprintnumbers,amsmath]{revtex4}

\usepackage[usenames]{color}
\usepackage{graphicx}
\usepackage{epsfig}
\usepackage{dcolumn}
\usepackage{bm}
\usepackage{latexsym}
\usepackage{amsfonts}
\usepackage{amssymb}

\newcommand{\beq}{\begin{equation}}
\newcommand{\eeq}{\end{equation}}
\newcommand{\beqa}{\begin{eqnarray}}
\newcommand{\eeqa}{\end{eqnarray}}
\newcommand{\bea}{\begin{array}}
\newcommand{\ena}{\end{array}}

\begin{document}
\title{Properties of global monopoles with an event horizon}
\author{Takashi Tamaki}
\email{tamaki@tap.scphys.kyoto-u.ac.jp}
\affiliation{Department of Physics, Kyoto University, 
606-8501, Japan}
\author{Nobuyuki Sakai}
\email{sakai@ke-sci.kj.yamagata-u.ac.jp}
\affiliation{Faculty of Education, Yamagata University, 
Yamagata 990-8560, Japan}
\address{Graduate School of Science and Engineering, Yamagata 
University, Yamagata 990-8560, Japan}
\address{Osservatorio Astronomico di Roma, Via Frascati, 00040 Monte 
Porzio Catone, Italia}

\date{\today}
\begin{abstract}
We investigate the properties of global monopoles with an event horizon. 
We find that there is an unstable circular orbit even if a 
particle does not have an angular momentum when the core mass is negative. 
We also obtain the asymptotic form of solutions 
when the event horizon is much larger than the core radius of the monopole, 
and discuss if they could be a model of galactic halos.
\end{abstract}

\preprint{KUNS-1862}

\maketitle

\section{Introduction}
In unified theories, various kinds of topological defects have 
been predicted and may appear in cosmological phase transitions in the early universe. 
Therefore it is important to investigate such defects both theoretically and
observationally. Concerning global monopoles, important facts were found recently: 
(i) Solutions with an event horizon exist \cite{Liebling,Sudarsky,Maison}. 
(ii) Regular global monopoles coupled nonminimally to gravity have a stable circular 
orbit and 
may explain rotation curves in spiral galaxies, as shown by Nucamendi {\it et al.}
(hereafter, NSS) \cite{Nucamendi}. 

The result (i) is interesting since the static solutions of regular global monopoles
are always repulsive \cite{Harari}. Moreover, black hole solutions are stable though 
topological charge is lost in the strict sense \cite{Watabe} and there are solutions 
with zero mass which are somewhat pathological \cite{Sudarsky}. Thus, we need to 
understand their properties. In particular, it would be interesting to investigate
a particle motion around the horizon. This is one of our main concerns in this paper.

Possibility (ii) shows that global monopoles can be locally attractive in 
the nonminimally coupled theory of gravity. 
Although there have been many attempts to explain rotation curves in spiral galaxies,
there is no definite one at present. Among solitonic objects, global monopoles have
a remarkable property that energy density decreases with the distance
$r^{-2}$ \cite{Barriola}, which may be desirable to explain flatness of rotation
curves. To remove the unprefereble repulsive property of global monopoles, NSS
introduced the nonminimal coupling and succeeded to obtain locally attractive
solutions.

Both (i) and (ii) taking into consideration, we notice a possibility that 
global monopoles with an event horizon can explain rotation curves
since they would be attractive. There are several advantages in this model compared
with the NSS model. First, we need not require the nonminimal coupling,
which are constrained astrophysically \cite{Will}. 
Secondly, they would also be model black holes in the central galaxies. 
Thirdly, the core mass can be chosen to be astronomically large, contrary to the NSS 
model, where the core mass is necessarily microscopic.
Therefore it is important to study the properties of such global monopoles, and 
discuss whether or not they can be a realistic candidate as galactic 
halos, taking astrophysical bounds into account \cite{Hiscock} .

This paper is written as follows. In Sec.~II, we explain our model and basic equations. 
In Sec.~III, we investigate global monopoles with an event 
horizon in two situations separately: In Sec.~III-A, we consider the case where the 
size of 
event horizon is comparable to the core radius of the monopole to compare with 
regular monopoles. In Sec.~III-B, we consider the case where the size of an event 
horizon is 
astrophysically large. In Sec.~IV, we denote concluding remarks and discuss problems 
concerning the restriction from observation. 

\section{Basic Equations for Numerical Analysis}
We begin with the action 
\beqa
S  =  \displaystyle\int d^{4}x \sqrt{-g}\left[
\frac{R}{16\pi G}-\frac{(\nabla \Phi^{a} )^{2}}{2}
-\frac{\lambda}{4}(\Phi^{a}\Phi^{a}-v^{2})^{2}
\right],     \label{B3} 
\eeqa
where $G$ and $\Phi^{a}$ are the gravitational constant and 
the real triplet Higgs field, respectively. 
The theoretical parameters $v$ and $\lambda$ are the 
vacuum expectation value  and the self-coupling 
constant of the Higgs field, respectively. 

We assume that a space-time is static and spherically 
symmetric, in which the metric is written as 
\beqa
ds^{2}=-f(r)e^{-2\delta (r)}dt^{2}+
f(r)^{-1}dr^{2}+r^{2}d\Omega^{2}, 
\label{metric}
\eeqa
where $f(r):=1-2Gm(r)/r$. 
We adopt the hedgehog ansatz given by 
\beqa
\Phi^{a}  =  v \mbox{\boldmath $r$}^{a} h(r) ,
\eeqa
where $ \mbox{\boldmath $r$}^{a} $ is a unit radial vector. 

Under the above assumptions, the basic equations are
\beqa
\hspace{-5mm}\bar{m}'&=&\bar{r}^{2}\bar{v}^{2}\left[
\frac{f}{2}(h')^{2}+U\right]\ ,  
\label{m} \\
\hspace{-5mm}\delta '&=&-\bar{v}^{2}\bar{r}(h')^{2}\ , 
\label{del} \\
\hspace{-5mm}h''&=&-\frac{h'}{\bar{r}}+\frac{1}{f}\left[
h(h^{2}-1)+\frac{2h}{\bar{r}^{2}}+2\bar{r}\bar{v}^{2}h'U
-\frac{h'}{\bar{r}}\right]\ ,
\label{h}
\eeqa
where $'=d/d\bar{r}$ and
\beqa
U&=&\frac{(h^{2}-1)^{2}}{4}+\frac{h^{2}}{\bar{r}^{2}}\ .
\eeqa
We have introduced the following dimensionless variables:
\beqa
\bar{r}=v\sqrt{\lambda}r, \;\; \bar{m}=Gv\sqrt{\lambda}m, 
\;\; \bar{v}=v\sqrt{4\pi G}\ . 
\eeqa
We assume the regular event horizon at $r=r_{H}$. 
\beqa
\bar{m}_{H}&=&\frac{\bar{r}_{H}}{2},\;\; \delta_H< \infty ,
\label{mth}  \\
h_{H}'&=&\frac{h_{H}[2+\bar{r}_{H}^{2}(h_{H}^{2}-1)]}{
\bar{r}_{H}(1-2\bar{r}_{H}^{2}\bar{v}^{2}U_{H})} .
\label{hth}
\eeqa
The variables with subscript $H$ are evaluated 
at the horizon. 

We introduce the variable 
\beqa
\bar{m}_{core}(\bar{r}):=\bar{m}(\bar{r})-\bar{v}^{2}\bar{r} . 
\label{mcore}
\eeqa
and assume the boundary conditions at spatial infinity as
\beqa
\bar{m}_{core}(\infty)=: \bar{M}=const.,\ \ \delta (\infty)=0,\ \  
h(\infty)=1\ , 
\label{atinf}
\eeqa
which means that the space-time is asymptotically ``flat" with deficit angle. 
Here $\bar{M}$ corresponds to a core mass of the monopole, which determines a particle
motion we will see below. We will obtain the black hole solutions numerically by
solving Eqs.(\ref{m}) - (\ref{h})
iteratively with the boundary conditions (\ref{mth}), (\ref{hth}) and (\ref{atinf}). 

\section{Properties}

Space-time structure of global monopole black holes depends on the 
expectation value of the Higgs field \cite{Liebling}. 
For $\bar{v}^{2}<1/2$, there is a solution with asymptotically ``flat" 
space-time. We concentrate on this realistic case.

\subsection{small horizon}
Typically, it is supposed that a global monopole has a nontrivial structure in the
core $r\lesssim r_{core}:=2/v\sqrt{\lambda}$ \cite{Harari}, while the field is
almost constant, $h\cong 1$, outside the core $r\gtrsim r_{core}$. Therefore, we 
expect that a monopole black hole with a small horizon $r_H\cong r_{core}$
(i.e., $\bar{r}_{H}\cong 1$) may have new properties.
 
\begin{figure}[htbp]
\psfig{file=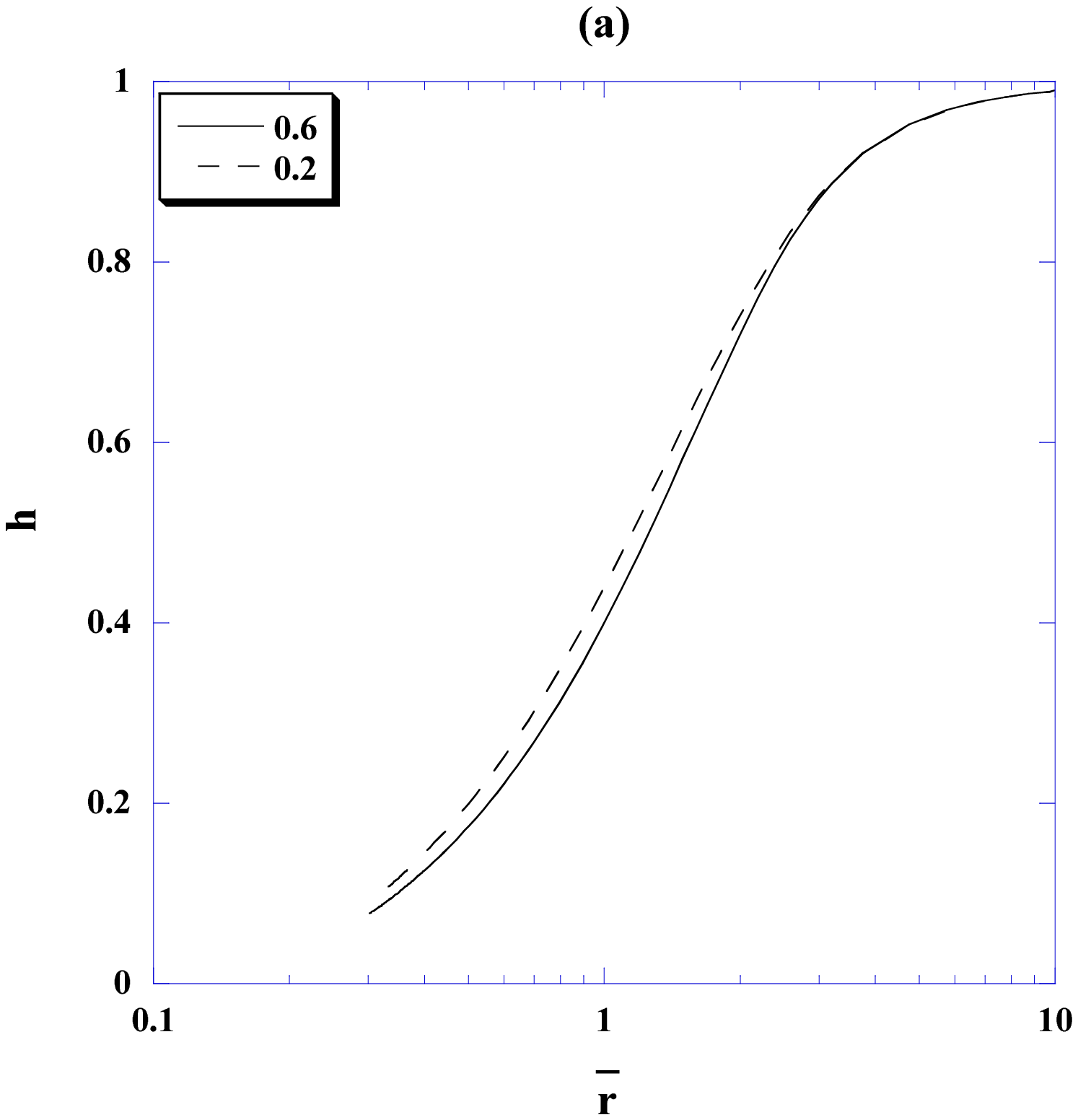,width=3.5in}
\psfig{file=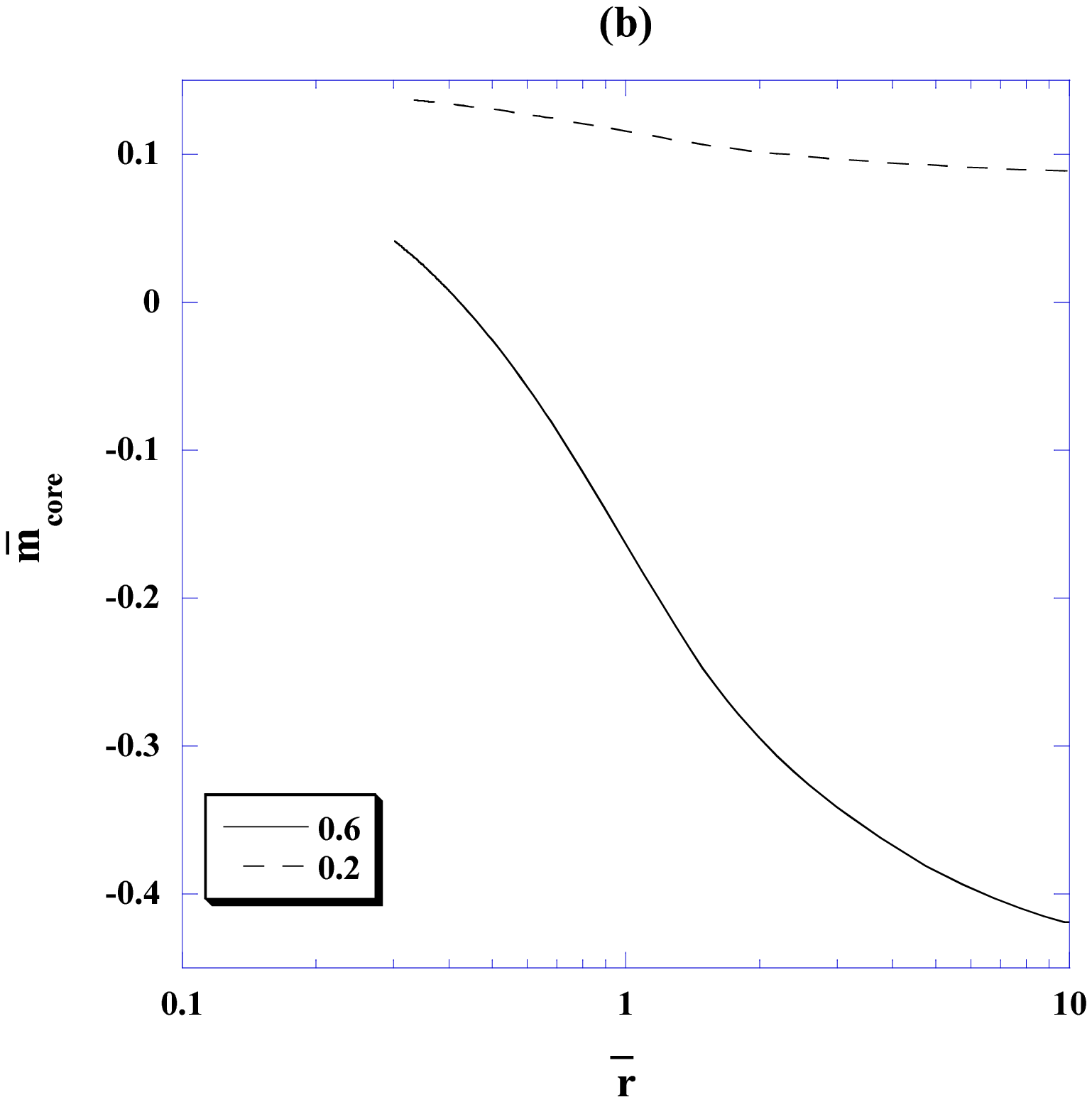,width=3.5in}
\caption{The behavior of (a) $h$ (b) $\bar{m}_{core}$ for $\bar{r}_{H}=0.3$ 
and $\bar{v}=0.6$, $0.2$. \label{r-field} }
\end{figure}

First, we show field distributions of black hole solutions
in Figs.~\ref{r-field} (a) $\bar{r}$-$h$ and (b) $\bar{r}$-$\bar{m}_{core}$. 
We choose $\bar{v}=0.6$, $0.2$ and $\bar{r}_{H}=0.3$. In Fig.~\ref{r-field} (a), 
we find that the Higgs field has a nontrivial structure extending to $\sim 10$ 
and it does not depend on the expectation value 
of the Higgs field, which is important in the later analysis.
In this solution, $h$ increases monotonically with $\bar{r}$.
Although the solutions where $h$ is not monotonic exist \cite{Maison}, here we only
consider monotonic solutions for simplicity.
Fig.~\ref{r-field} (b) shows that $\bar{m}_{core}$ for $\bar{v}=0.6$ decreases with 
$\bar{r}$ much faster than that for $\bar{v}=0.2$. This is natural because of the
factor $\bar{v}^{2}\bar{r}$ in (\ref{mcore}). The important point is that
$\bar{m}_{core}$ becomes 
negative for $\bar{v}=0.6$ and positive for $\bar{v}=0.2$ in the asymptotic region.
As we will discuss below, the sign of $\bar{m}_{core}$ at large $r$ (i.e., $\bar M$)
determines the qualitative behavior of particle motions around the monopole.

Let us consider the geodesic equation of a test particle in the 
equatorial plane. In our coordinate system (\ref{metric}), this is expressed as 
\beqa
\frac{E^{2}}{2}=\frac{\dot{r}^{2}}{2}e^{-2\delta}+V_{eff}, 
\label{geodesics}
\eeqa
where $\dot{}=d/dt$ and $E$ is the energy of the particle per unit rest mass. 
$V_{eff}$ is defined as 
\beqa
V_{eff}:=e^{-2\delta}\left( 1-\frac{2Gm}{r} \right) \left(1+
\frac{L^{2}}{r^{2}}\right) , 
\label{Veff}
\eeqa
where $L$ is the angular momentum of the particle per unit rest mass. 

\begin{figure}[htbp]
\psfig{file=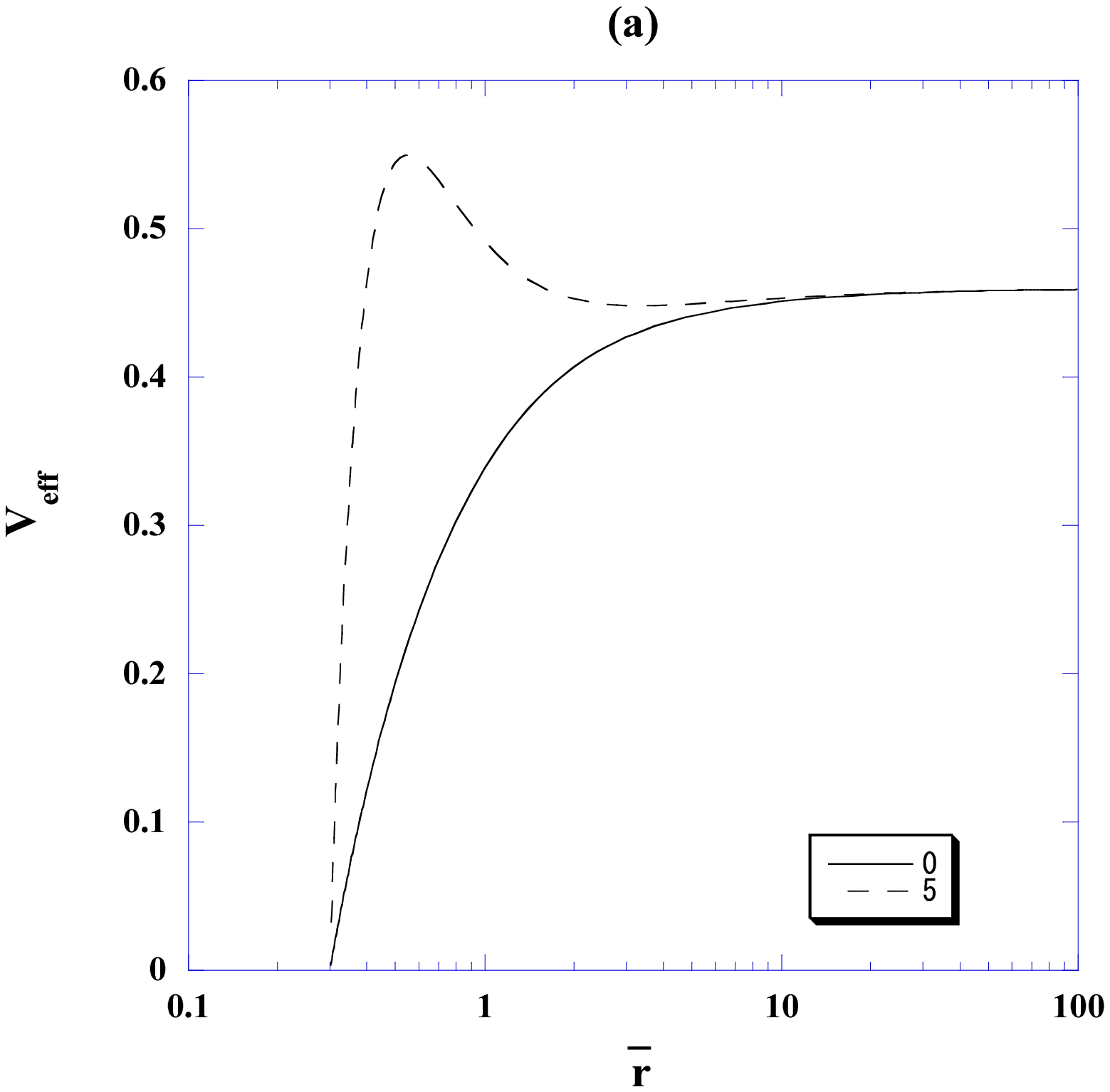,width=3.5in}
\psfig{file=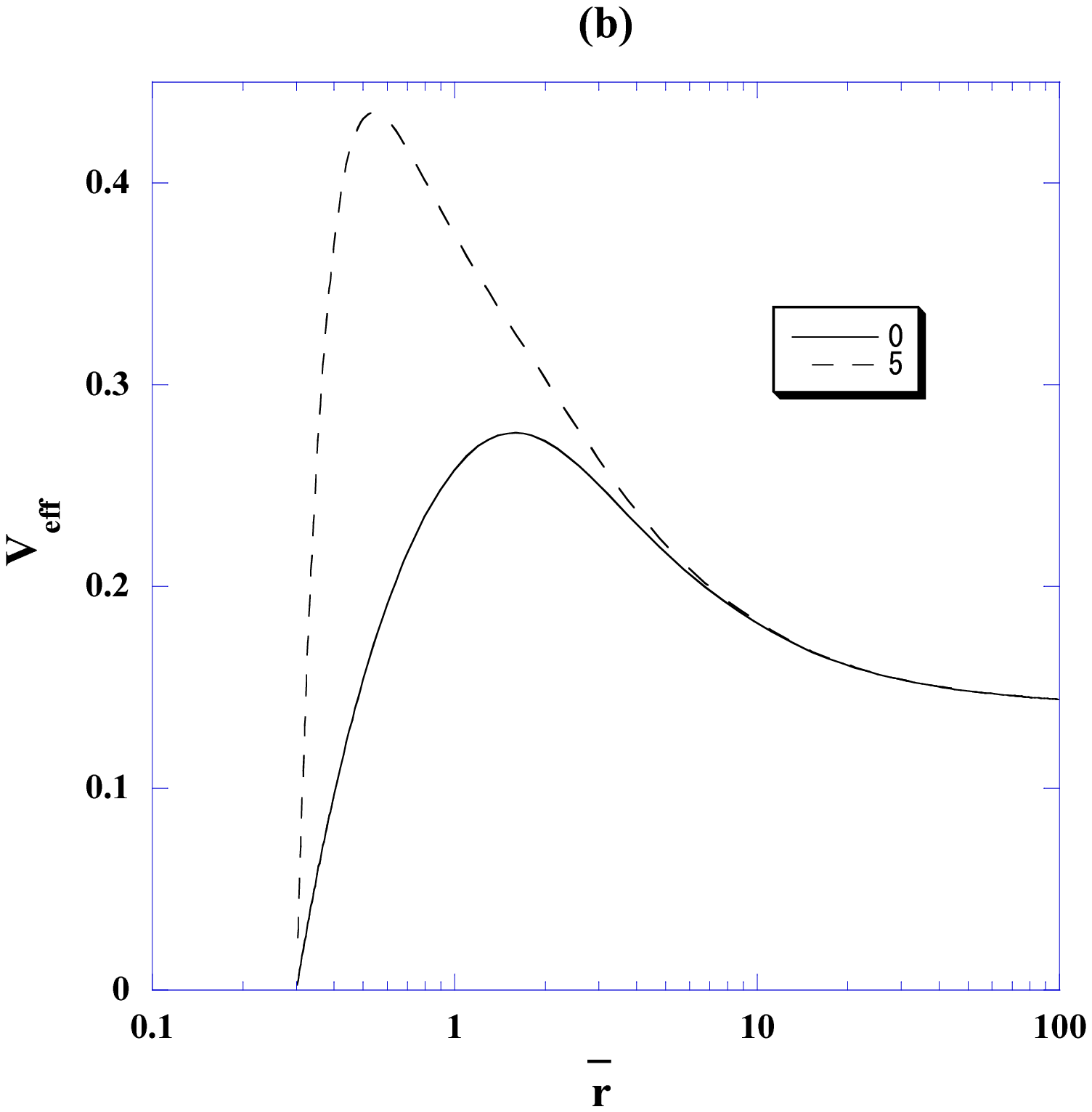,width=3.5in}
\caption{The behavior of $V_{eff}$ for $\bar{r}_{H}=0.3$, 
and $(L/r_{H})^{2}=0$, $5$. (a) $\bar{v}=0.2$ and (b) $\bar{v}=0.6$. 
\label{r-V} }
\end{figure}

We show the effective potential $V_{eff}$ for $\bar{r}_{H}=0.3$ 
in Figs.~\ref{r-V} (a) $\bar{v}=0.2$ and (b) $\bar{v}=0.6$. 
The angular momentum is chosen as $(L/r_{H})^{2}=0$ and $5$. 
For $(L/r_{H})^{2}=5$, there is no potential minimum for $\bar{v}=0.6$, while there
is for $\bar{v}=0.2$. 
For $(L/r_{H})^{2}=0$, there is a potential {\it maximum} for $\bar{v}=0.6$, while
there is not for $\bar{v}=0.2$. 
While the properties of the monopole black hole with $\bar{v}=0.2$ are essentially
the same as those of the Schwarzschild black hole, larger $\bar{v}$ changes properties
qualitatively.
These different properties are determined by the sign of $\bar{M}$. 

To evaluate the potential minimum, we substitute the asymptotic form
$\delta \to 0$ and $\bar{m}\to \bar{v}^{2}\bar{r}+\bar{M}$ into Eq.~(\ref{Veff}).
Then, $dV_{eff}/d\bar{r}=0$ is satisfied at the positions
\beqa
\bar{r}_{\pm}=\frac{\bar{L}^{2}(1-2\bar{v}^{2})\pm\sqrt{\bar{L}^{4}(1-
2\bar{v}^{2})^{2}-
12\bar{L}^{2}\bar{M}^{2} }}{2\bar{M}}\ , 
\label{minimum}
\eeqa
where $\bar{L}$ is defined as $\bar{L}:=v\sqrt{\lambda}L$.
We find $\bar{r}_{\pm}<0$ when $\bar{L}^{2}(1-2\bar{v}^{2})^{2}>12\bar{M}^{2}$ 
and $\bar{M}<0$. Therefore, there is no potential minimum even if a test particle 
has large angular momentum.

On the other hand, the potential maximum cannot be evaluated from the asymptotic form
of the solution because it is near the horizon if it exists, as shown by Fig.\ 2. 
However, we can discuss its existence as follows. 
If $\bar{M}<0$, $V_{eff}$ decreases with $\bar{r}$ and approaches $1-2\bar{v}^{2}$ 
Asymptotically; since $V_{eff}=0$ at the horizon, there is at 
least one local maximum at some $\bar{r}$. 
If $\bar{M}>0$, on the other hand, $V_{eff}$ increases with $\bar{r}$ 
and approaches $1-2\bar{v}^{2}$ asymptotically; whether or not the potential 
maximum appear depends on $\bar{L}$ as in the Schwarzschild black hole. 
Thus, the sign of $\bar{M}$ is important to determine the particle motion around the
black hole.

Such a convex form of the potential for a $L=0$ particle motion is characteristic
of a global monopole black hole, and does not appear neither in a Schwarzschild black
hole nor in a regular global monopole. In the case of a regular global monopole, the whole
space-time is repulsive \cite{Harari}, which means that $V_{eff}$ decreases
monotonically.

\begin{figure}[htbp]
\psfig{file=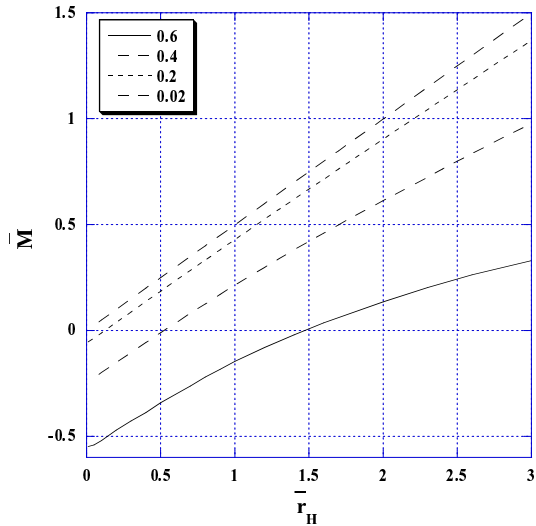,width=3.5in}
\caption{The relation between $\bar{r}_{H}$ and $\bar{M}$ 
for various $\bar{v}$. \label{rh-Mcore} }
\end{figure}

Fig.~\ref{rh-Mcore} shows the relation between $\bar{r}_{H}$ and $\bar{M}$ for
various $\bar{v}$. $\bar{M}$ is negative in the limit $\bar{r}_{H}\to 0$ 
as it is expected from the regular solutions. As $\bar{v}$ increases, the region of
$\bar{M}<0$ extends to larger $\bar{r}_{H}$. As it was pointed out in
Ref.~\cite{Sudarsky}, there are solutions with $\bar{M}=0$. In this
case, a potential minimum 
$\bar{r}_{-}$ goes to infinity and a test particle cannot feel a black hole. 

\subsection{large horizon}
NSS argued that global monopoles with the nonminimally coupled gravity may explain 
rotation curves in spiral galaxies \cite{Nucamendi}. However, the nonminimally
coupled gravity has been 
constrained astrophysically. Moreover, in the NSS model the bound orbits exist only
in the microscopic region $r\lesssim r_{core}$.
Thus, it is desirable to seek for other possibilities. 

On the other hand, Wetterich discussed the possibility that a massless scalar field
may explain rotation curves of galactic halos \cite{Wetterich}. 
In his solutions, however, physical boundary conditions were not taken into account. 
If we assume regularity at the center or an event horizon of a black hole, only a
trivial solution remains. Nonexistence of nontrivial black hole solutions are
guaranteed by no hair theorem \cite{Bekenstein}. In this sense, his model is also 
unrealistic. 

Then, we turn to a global monopole with an event horizon. This model is free from
the above difficulties existing in the NSS model. Furthermore, the existence of an 
event horizon is realistic 
because massive black holes are observed in the central regions of galaxies.

Let us consider astrophysical bounds from the mass density of 
monopoles in the universe at first. If we demand that mass density of monopoles 
should be less than 10 times of critical density \cite{Hiscock}, we have 
\beqa
n<10^{-3}\left(\frac{10^{16}{\rm GeV}}{v}\right){\rm Mpc}^{-3}\ ,
\label{number}
\eeqa
where $n$ is the number density of monopoles. For definiteness, we choose 
$\bar{v}=0.2\times 10^{-4}$ and discuss the case where the event horizon is
cosmological size. 

\begin{figure}[htbp]
\psfig{file=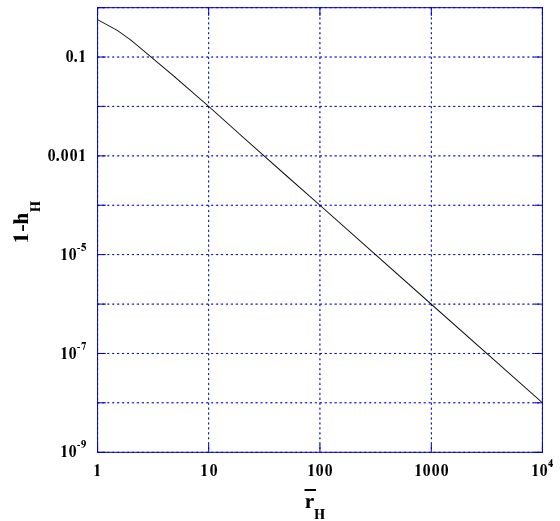,width=3.5in}
\caption{The relation between $\bar{r}_{H}$ and $(1-h_{H})$ 
for $\bar{v}=0.2\times 10^{-4}$. \label{rh-(1-hh)} }
\end{figure}

To see the structure of the Higgs field for $\bar{r}_{H}\gg 1$, we plot
$r_{H}$-$(1-h_{H})$ in Fig.~\ref{rh-(1-hh)}. We find $h_H\cong1$ and
$(1-h_{H})\cong 1/r_{H}^{2}$. We can check this relation analytically as follows.
In the asymptotic region, supposing the asymptotic form,
\beqa
h=1+\sum_{n=1}^{\infty}C_{n}r^{-n}\ ,
\eeqa
we have $C_{1}=0$ and $C_{2}=-1$, which are consistent with our numerical results
above.

\begin{figure}[htbp]
\psfig{file=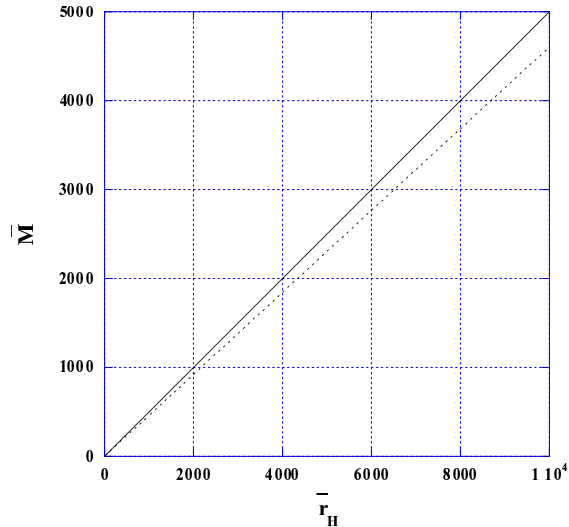,width=3.5in}
\caption{The relation between $\bar{r}_{H}$ and $\bar{M}$ 
for $\bar{v}=0.2\times 10^{-4}$ and $\bar{v}=0.2$ by a solid line and by 
a dotted line, respectively. \label{r-mcorel} }
\end{figure}

Because of this asymptotic behavior, $\bar{M}$ can be estimated by substituting $h=1$ 
into Eq.~(\ref{m}). Then, we have the asymptotic relation,
\beqa
\bar{M}=\frac{\bar{r}_{H}}{2}(1-2\bar{v}^{2})\ .
\label{large}
\eeqa
We show the relation $\bar{r}_{H}$-$\bar{M}$ in Fig.~\ref{r-mcorel} for 
$\bar{v}=0.2\times 10^{-4}$ and $\bar{v}=0.2$, which confirms the relation
(\ref{large}). Actually, Fig.~\ref{rh-Mcore} shows that this approximation
is fairly good even for $\bar{r}_{H}\sim 1$ when $\bar{v}$ is small. 

Let us consider particle motions. Setting $h=1$, we have
\beqa
V_{eff}\sim \left(1-2\bar{v}^{2}+\frac{2\bar{M}}{\bar{r}}\right)\left(1+
\frac{\bar{L}^{2}}{\bar{r}^{2}}\right)\ .
\eeqa
Since $\bar{v}$ is small as constrained by (\ref{number}), we find that the particle
motion is practically the same as that in Schwarzschild black hole. 
As long as we consider an astronomical-sized event horizon, the effect of
scalar fields on particle motions is negligible. 

\section{Conclusion and discussion}

We investigated properties of global monopoles with an event horizon 
and revealed interesting features which had not been known so far. 
The main features of test particle motions are determined by the sign of the
core mass; if it is negative and if the event horizon is as small as the core radius,
there is an unstable circular orbit even for a particle with zero angular momentum.
We also found the asymptotic form of the solutions when the event horizon is much
larger than the core radius; the qualitative features of the monopole black hole is
the same as that of the Schwarzschild black hole.

Although our model does not explain observed rotation curves very well, we obtain some
lessons here. Our results indicates that massive scalar fields would encounter
with the same difficulty as in our model. A typical mass scale of particle physics
is so large that it generally contributes only to microscopic structure, whose size is of order 
of the inverse of the mass. In this sense, a massless scalar field considered by Wetterich 
\cite{Wetterich} might be useful. Although his model itself does not satisfy physical boundary 
conditions neither of a black hole nor of a regular solution, interaction with matter may be a 
key ingredient to solve this problem. Including this possibility, we also want to consider other 
soliton-like models such as boson-fermion stars in future \cite{Henriques}. 

\section*{ACKNOWLEDGEMENTS}

N.S.\ thanks Luca Amendola for discussions and his hospitality at 
Osservatorio Astronomico di Roma. 
This work was supported in part by Grant-in-Aid for Scientific Research Fund of the
Ministry of Education, Science, Culture and Technology of Japan, 2003, No.\ 154568
(T.T.) and No.\ 00267402 (N.S.).  This work was also supported in part by a
Grant-in-Aid for the 21st Century COE ``Center for Diversity and Universality in 
Physics", and by the exchanging-researcher project between Japan Society for the
Promotion of Science and National Research Council of Italy, 2003.



\end{document}